\documentstyle[aps,preprint]{revtex}
\begin{document}
\draft
\title{\bf Filtering number states of the vibrational motion of an ion}
\author{H. Moya-Cessa\footnote{On leave from: INAOE, Coordinaci\'on de Optica,
Apdo. Postal 51 y 216, 72000 Puebla, Pue., Mexico} and P. Tombesi}
\address{Via Madonna delle Carceri, Dipartimento di Matematica e Fisica,
Universit\`a di Camerino and INFM unit\`a Camerino, I-62032
Camerino (MC), Italy}
\date{\today}
\maketitle
\begin{abstract}
We propose a scheme to generate number states, and specific superpositions
of them, of the vibrational motion of a trapped 
ion via laser excitation of two vibronic transitions. Without detrimental effects
on the internal state preparation, the ion can be initially in thermal motion 
and undergo spontaneous emission. In particular, robust to
noise qubits ($\alpha|0\rangle+\beta|1\rangle$ states) can be generated with arbitrary
amplitudes.
\end{abstract}
Recently major efforts have been directed towards the generation of 
non-classical states of light \cite{kurizki,dariano}
and vibrational motion of ions \cite{vogeli,vogelii,moyai}.
In particular, number states and specific superpositions of them are of importance 
for a number of different applications ranging from spectroscopy to fundamental
tests of quantum mechanics.
Methods to engineer quantum states of light have already been proposed which
are mainly based on conditional measurements \cite{kurizki,moyai}. 
There have been proposed methods
to restore \cite{mauro}  and protect \cite{vitali} states  
and also to engineer not a state itself but the Hamiltonian \cite{vogeliii}.
Here we propose to generate number states and given superpositions of them, 
with controllable spacing, by bicromatic excitation of an ion which can be 
realized with present available technology. Our scheme is based on 
proposals by
de Matos Filho and Vogel where they show that Schr\"odinger cats \cite{vogeli} and 
nonlinear coherent states \cite{vogelii} emerge from the long time
dynamics, when the ion decouples from the laser field and remains
in a pure state. In this contribution, by using the same scheme
as in \cite{vogeli,vogelii}  
we exploit the fact that, by manipulating
the Lamb-Dicke parameters (LDPs) of both fields, under certain
conditions we can generate superpositions
of spaced number states below a given definite number, above it, or generate a single 
number state, therefore "filtering" (choosing) number states. In particular, robust to
noise qubits \cite{ekert} ($\alpha|0\rangle+\beta|1\rangle$ states) can be generated with arbitrary
amplitudes.

We consider the Hamiltonian of a single ion trapped in a harmonic 
potential in interaction with 
laser  light in the rotating wave approximation, which reads 

\begin{equation}
\hat{H} = \hbar\nu\hat{a}^\dagger \hat{a} + \hbar\omega_{21}\hat{A}_{22}
+\hbar[\lambda E^{(-)}(\hat{x},t)\hat{A}_{12}+ H.c.],
\label{1}
\end{equation}
where $\hat{a}$  and  $\hat{A}_{ab}$ 
are the annihilation operator of  a quantum of the ionic vibrational motion and
the electronic (two-level) flip operator for the $|b\rangle \rightarrow |a\rangle$ 
transition of frequency $\omega_{21}$,
respectively. $\nu$ is the trap frequency, $\lambda$  the electronic coupling 
matrix element, and $E^{(-)}(\hat{x},t)$ the negative part of the classical electric 
field of the driving field.

We assume the ion driven by two laser fields, the first tuned to the $j$th 
lower sideband and the second tuned to the $m$th lower sideband, therefore we may write
$ E^{(-)}(\hat{x},t)$ as

\begin{equation}
E^{(-)}(\hat{x},t)=
E_{j}e^{-i(k_{j}\hat{x}-\omega_{21}+j\nu)t}+
E_{m}e^{-i(k_m\hat{x}-\omega_{21}+m\nu)t},
\label{2}
\end{equation}
where, if $m=0$ it would
correspond to the driving field being on 
resonance with the electronic transition, and $\hat{x}$ may be written as
\begin{equation}
k_s\hat{x}=\eta_s(\hat{a}+\hat{a}^\dagger),
\label{3}
\end{equation}
where $k_s$ are the wave vectors of the driving fields and
\begin{equation}
\eta_s=2\pi\frac{\sqrt{\langle 0|\Delta\hat{x}^2|0\rangle}}{\lambda_s}
\end{equation}
are the LDPs with $s=j,m$.

It has been noted recently, that approximations based on this
parameter can lead to divergencies in the mean motional excitation \cite{vogeliv}. A way of 
linearizing the Hamiltonian (\ref{1}) which accounts for all orders of the Lamb-Dicke 
parameter and any sideband regime has been proposed in \cite{moyaii}.

In the resolved sideband limit, the vibrational frequency $\nu$ is much larger than 
other characteristic frequencies and the interaction of the ion with the two lasers
can be treated separately, using a nonlinear Hamiltonian \cite{vogeli}.  The Hamiltonian
(\ref{1}) in the interaction picture can then be written as
\begin{equation}
\hat{H}_I=\hbar\hat{A}_{21} \left[\Omega_j e^{-\eta_{j}^2/2}
\frac{\hat{n}!}{(\hat{n}+j)!}L_{\hat{n}}^{(j)}(\eta_{j}^2)\hat{a}^j
+\Omega_m e^{-\eta_{m}^2/2}
\frac{\hat{n}!}{(\hat{n}+m)!}L_{\hat{n}}^{(m)}(\eta_{m}^2)\hat{a}^m
\right]+H.c.,
\label{4}
\end{equation}
where $L_{\hat{n}}^{(k)}(\eta_{k}^2)$ are the operator-valued associated Laguerre
polynomials and the $\Omega$'s are the Rabi frequencies and $\hat{n}=
\hat{a}^\dagger\hat{a}$.
The master equation which describes this system can be written as
\begin{eqnarray}
         \frac{ \partial \hat{\rho}}{\partial t} = -\frac{i}{\hbar} [\hat{H}_I,\hat{\rho}] +
         \frac{\Gamma}{2} \left(
           2\hat{A}_{12} \hat{\tilde{\rho}} \hat{A}_{21}
             - \hat{A}_{22} \hat{\rho} - \hat{\rho} \hat{A}_{22} \right) 
         \label{5}
\end{eqnarray} 
where the last term describes spontaneous emission with energy relaxation rate 
$\Gamma$, and
\begin{equation}
 \hat{\tilde{\rho}}=\frac{1}{2}\int^1_{-1}ds W(s)e^{is\eta_E\hat{x}}\hat{\rho}
e^{-is\eta_E\hat{x}},
\label{6}
\end{equation}
accounts for changes of the vibrational energy because of spontaneus emission.
Here $\eta_E$ is the LDP corresponding to the field (\ref{2}) and
$W(s)$ is the angular distribution of spontaneus emission \cite{vogeli,vogelii}.

The steady-state solution to Eq. (\ref{5}) is obtained by setting $\partial 
\hat{\rho}/\partial t=0$ and may be written as \cite{vogeli,vogelii}
\begin{equation}
 \hat{\rho}_s= |1\rangle|\psi_s\rangle\langle\psi_s|\langle1|,
\label{7}
\end{equation}
where $|1\rangle$ is the electronic ground state and $|\psi_s\rangle$ is the 
vibrational steady-state of the ion, given by
\begin{equation}
\left(\Omega_j e^{-\eta_{j}^2/2}
\frac{\hat{n}!}{(\hat{n}+j)!}L_{\hat{n}}^{(j)}(\eta_{j}^2)\hat{a}^j
+\Omega_m e^{-\eta_{k}^2/2}
\frac{\hat{n}!}{(\hat{n}+m)!}L_{\hat{n}}^{(m)}(\eta_{m}^2)\hat{a}^m
\right)
|\psi_s\rangle=0.
\label{8}
\end{equation}

For simplicity, we will concentrate in the $j=1$ and $m=0$ case (single number state spacing)
for which Eq. (\ref{8}) is written as
\begin{equation}
\left(\Omega_1 e^{-\eta_{1}^2/2}
\frac{L_{\hat{n}}^{(1)}(\eta_{1}^2)}{\hat{n}+1}\hat{a}
+\Omega_0 e^{-\eta_{0}^2/2}
L_{\hat{n}}(\eta_{0}^2)\right)
|\psi_s\rangle=0.
\label{9}
\end{equation}
Note that $\hat{H}_I|1\rangle|\psi_s\rangle=0$ so that ion and laser have stopped to interact,
which occurs when the ion stops to fluoresce. For the $j=1$ and $k=0$ case, and 
assuming $L_k^{(1)}(\eta_1^2)\neq 0$ and $L_k(\eta_0^2)\neq 0$ for all $k$, one 
generates nonlinear coherent states \cite{vogelii}. However, by setting a value to 
one of the LDPs such that, for instance, 
\begin{equation}
L_q(\eta_0^2)=0,
\label{10}
\end{equation}
for some integer $q$ (but $L_k^{(1)}(\eta_1^2)\neq 0$ for all $k$),
we obtain that, by writing $|\psi_s\rangle$ in the number state representation,
\begin{equation}
|\psi_s(\eta_0)\rangle=\frac{1}{N_0}\sum_{n=0}^qC^{(0)}_n|n\rangle,
\label{11}
\end{equation}
(the argument of $\psi_s$ denotes the condition we apply, i.e., in Eq (\ref{11}),
the condition is on $\eta_0$)
where 
\begin{equation}
C^{(0)}_n=\left(-\frac{\Omega_0 e^{-\eta_0^2/2}}{\Omega_1 e^{-\eta_1^2/2}}\right)^n
(n!)^{1/2}\prod_{m=0}^{n-1}\frac{L_m(\eta_0^2)}{L_m^{(1)}(\eta_1^2)}, \ \ \ \ \
C^{(0)}_0=1,
\end{equation}
and
\begin{equation}
N_0^2=\sum_{n=0}^q|C_n^{(0)}|^2
\end{equation}
is the normalization constant. 

If, instead of condition (\ref{10}), we choose
\begin{equation}
L_p^{(1)}(\eta_1^2)=0,
\label{12}
\end{equation}
(but $L_k(\eta_0^2)\neq 0$ for all $k$) we obtain the wave function
\begin{equation}
|\psi_s(\eta_1)\rangle=\frac{1}{N_1}\sum_{n=p+1}^\infty C_n^{(1)}|n\rangle,
\label{13}
\end{equation}
where now,
\begin{equation}
C^{(1)}_n=\left(-\frac{\Omega_0 e^{-\eta_0^2/2}}{\Omega_1 e^{-\eta_1^2/2}}\right)^{n-p-1}
\sqrt{\frac{n!}{(p+1)!}}\prod_{m=p+1}^{n-1}\frac{L_m(\eta_0^2)}{L_m^{(1)}(\eta_1^2)}, \ \ \ \ \
C^{(1)}_{p+1}=1,
\end{equation}
and
\begin{equation}
N_1^2=\sum_{n=p+1}^\infty|C_n^{(1)}|^2.
\end{equation}
Combining both conditions (\ref{10}) and (\ref{12}), one would obtain (for $q>p$)
\begin{equation}
|\psi_s(\eta_0,\eta_1)\rangle=\frac{1}{N_{01}}\sum_{n=p+1}^q C^{(1)}_n|n\rangle,
\label{14}
\end{equation}
with 
\begin{equation}
N_{01}^2=\sum_{n=p+1}^q|C_n^{(1)}|^2.
\end{equation}
In this way, by setting the conditions (\ref{10}), (\ref{12}) or
both, we can engineer states in the following three zones of the Hilbert space, 
(a)  from $|0\rangle$ to $|q\rangle$, 
(b) from $|p+1\rangle$ to $|\infty\rangle$, or
(c) from $|p+1\rangle$ to $|q\rangle$. 
In the later case, 
by setting $q=p+1$
generation of the number state $|q\rangle$ is achieved.

We should remark, that by selecting further apart sidebands one would obtain
a different spacing in Eqs (\ref{11}), (\ref{13}) and (\ref{14}). For instance,
by choosing $j=2$ and $k=0$ one would obtain only even or odd number states in those 
equations (depending in this case on initial conditions and $W(s)$, the angular
distribution of spontaneus emission). 
Also, it should be noticed that one can use the parameters $j=m+1$ and $k=m$
(with $m \neq 0$)
(in the single number state spacing case) to extend the possibilities of choosing
LDPs. LDPs of the order of one (or less)
are needed [for
conditions (\ref{10}) and (\ref{12})] which can be achieved by varying the geometry 
of the lasers. For example, by setting $\eta_0=1$, we have $L_1(\eta^2_0=1)=0$, and
therefore we obtain the qubit
\begin{equation}
|\psi_s(\eta_0=1)\rangle=\frac{1}
{\sqrt{1+|\frac{\Omega_0}{\Omega_1}|^2 e^{\eta_1^2-1}}}
\left(|0\rangle-
\frac{\Omega_0 e^{-1/2}}{\Omega_1 e^{-\eta_1^2/2}}|1\rangle\right),
\label{15}
\end{equation}
where by changing the Rabi frequencies, one has control
of the amplitudes.

Finally, note that we could have also chosen to drive the $q$th upper sideband
instead of the $k$th lower sideband in equation (\ref{2})  with
bassically the same results.

In conclusion, we have shown that specific superpositions 
of number states with $k$th spacing emerge from the long time dynamics,
when laser and ion decouple and this stops fluorescing.  From Eq (\ref{14})
one notes that we can generate robust to noise qubits with controllable
amplitudes, which depend on the Rabi frequencies ($\Omega$'s). This can be easily 
done by varying the LDPs to fulfil  conditions 
(\ref{10}) and (\ref{12}). In this way, one has broad possibilities to "filter" (choose) 
which number states are present in the state to be generated.

H. M.-C. acknowledges W. Vogel and  S. Wallentowitz
for comments and M. Fortunato for discussions and proof-reading
of the manuscript. H. M.-C. would also like to thank support from the Mexican Consejo 
Nacional de Ciencia y Tecnolog\'ia (CONACYT).

%




%






%






%







\end{document}